\documentclass[aps,prl,twocolumn,groupedaddress,showpacs]{revtex4-1}
\usepackage{CJK}
\usepackage{graphicx,subfigure}
\usepackage{epstopdf}
\usepackage{amsmath}

\begin{document}

\title{Threshold Characteristics of Slow-Light Photonic Crystal Lasers}


\author{Weiqi Xue}
\email{wexu@fotonik.dtu.dk}
\author{Yi Yu}%
\author{Luisa Ottaviano}%
\author{Yaohui Chen}%
\author{Elizaveta Semenova}%
\author{Kresten Yvind}%
\author{Jesper Mork}%
\affiliation{DTU Fotonik, Department of Photonics Engineering, Technical University of Denmark, Building 343, DK-2800 Kongens Lyngby, Denmark}


\date{\today}

\begin{abstract}
The threshold properties of photonic crystal quantum dot lasers operating in the slow-light regime are investigated experimentally and theoretically. Measurements show that, in contrast to conventional lasers, the threshold gain attains a minimum value for a specific cavity length. The experimental results are explained by an analytical theory for the laser threshold that takes into account the effects of slow light and random disorder due to unavoidable fabrication imperfections. Longer lasers are found to operate deeper into the slow-light region, leading to a trade-off between slow-light induced reduction of the mirror loss and slow-light enhancement of disorder-induced losses. 
\end{abstract}


\maketitle
Slow light in photonic crystal (PhC) line-defect waveguides \cite{Baba2008} enhances the interaction between the propagating light wave and the material of the waveguide, and has enabled the demonstration of increased material nonlinearity \cite{Corcoran2009}, enhanced spontaneous emission into the propagating mode \cite{Noda2007, Lund2008}, and enhanced material gain \cite{Ek2014}. Such engineering of fundamental materials properties is important for the development of integrated photonic circuits, with applications in classical as well as quantum information technology. Microcavity lasers can be realized in the same PhC membrane structure by exploiting high-quality point-defect cavities and in the past decade significant progress was made \cite{Noda2006, Atlasov2009,Matsuo2010}, culminating in recent demonstrations of high-speed electrically pumped structures  \cite{Matsuo2013}. Such PhC lasers allow the exploration of new operation regimes, such as single emitter lasing \cite{Nomura2010} and ultra-high speed modulation \cite{Mork2014}. However, while it was shown that slow light in combination with random spatial disorder leads to very rich physics \cite{Hughes2005, Mazoyer2009, Patterson2009, LeThomas2009, Sapienza2010, Faolain2010, Liu2014}, the role of slow light on lasers realized using defect cavities has apparently not been systematically investigated. For the case of passive point-defect cavities, it is well known that disorder is an important factor limiting the quality factor \cite{Asano2006,Ramunno2009,Portalupi2011,Minkov2013} but the role of slow light in extended active cavities is not well understood.

In this paper, we report experimental results on PhC quantum dot lasers with variable cavity length and show that these attain a minimum threshold gain for a certain cavity length, in stark contrast to conventional lasers, where the threshold gain decreases monotonically with cavity length. We derive a rate equation including the effect of slow-light propagation and show that the experimental observations may be explained when taking into account disorder-induced losses. These results show that disorder may lead to fundamental limitations on the performance of nanostructured lasers, but the results also demonstrate a promising platform for investigating disorder effects in active structures, such as the competition between deterministic cavity modes and random modes formed by Anderson localization \cite{Liu2014}.

 \begin{figure}
 \includegraphics[width=5cm]{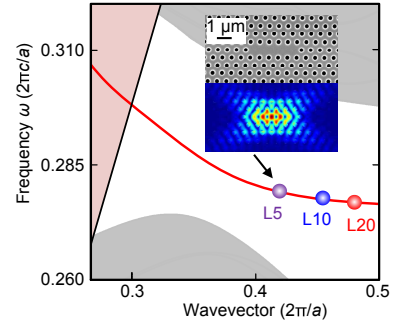}%
 \caption{\label{BandDiagram} Qualitative illustration of PhC banddiagram with TE-like even waveguide mode (solid line) and positions of the fundamental (M1) cavity mode for L5, L10, and L20 defect cavities (markers). Shaded areas indicate the light cone and non-guided modes of the structure. The inset shows a SEM image of an L5 PhC cavity laser and the simulated spatial mode profile of the electric field amplitude.}
 \end{figure}

The PhC cavity lasers investigated here are realized in a 250 nm thick InP air-embedded PhC membrane \cite{Xue2015} with three layers of quantum dots with densities of $\sim5.4\times10^{10}\ \text{cm}^{-2}$ \cite{Semenova2014}. The  PhC structure has a lattice constant of $a=438\ \text{nm}$ and an air-hole radius of $0.25a$. A so-called L$N$ cavity \cite{Okano2010} is formed by omitting $N$ air-holes in a W1 defect waveguide. Structures with different cavity lengths, i.e., L1-L20 are fabricated and characterized.  In Fig.~\ref{BandDiagram}, the red solid curve illustrates the dispersion of the even TE-like W1-waveguide mode and the corresponding  modes for L5, L10, and L20 cavities are represented by markers. The inset shows the SEM image of a fabricated L5 structure and the corresponding simulated mode profile of the norm of the field amplitude in the middle of the membrane. As the cavity length increases, the fundamental mode moves closer to the Brillouin zone (BZ) edge, where the group index strongly increases \cite{Baba2008}, thus allowing a systematic investigation of the role of slow light on the laser properties.

The samples are vertically pumped using a micro-photoluminescence set-up, with precise control of pump position and area, monitored by an infra-red camera. The laser emission is collected vertically by the same objective, with an NA of 0.65, and monitored by a spectrum analyzer. The transmission efficiency from the objective to the spectrum analyser is $\sim30\%$. All measurements are performed at room temperature. 

 \begin{figure}
 \includegraphics[width=8.0cm]{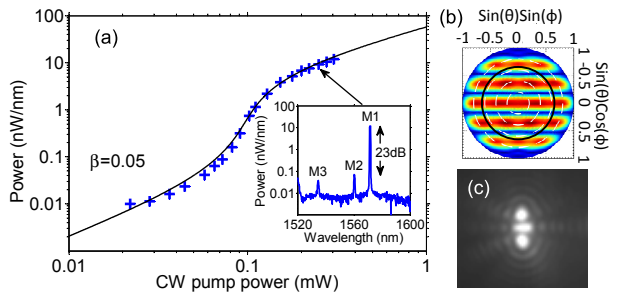}%
 \caption{\label{L9_Cavity} (a) Measured power of the fundamental mode M1 versus   pump power and corresponding theoretical fit (solid line) for an L9 laser. The inset shows the output spectrum far above threshold. (b) Simulated L9 far-field profile with circles indicating the collection area for different NAs, and (c) measured emission profile corresponding to black circle.}
 \end{figure}

Fig.~\ref{L9_Cavity}(a) shows a typical example of a measured light-out vs light-in characteristic, obtained by pumping an L9 cavity with the CW output from a 1480 nm diode. The pump power is estimated by multiplying the excitation power ($P_{exc}$) with the ratio ($\Gamma_{exc}$) of the PhC cavity area to the total pump area. The limited absorption efficiency of the QD layers was not taken into account. The inset of Fig.~\ref{L9_Cavity}(a) displays the emission spectrum, showing the dominance of the fundamental lasing mode, denoted M1, but also the presence of two higher order longitudinal cavity modes, M2 and M3. The side mode suppression increaes by further increase of the pump power. The solid line in Fig.~\ref{L9_Cavity}(a) shows a fit to a conventional laser rate equation model, see e.g. \cite{Matsuo2013}, leading to a laser threshold of $\sim90\ \mu \text{W}$ and a spontaneous emission factor of $\beta=0.05$. This $\beta$-factor is about an order of magnitude smaller than values reported in \cite{Matsuo2013}, but enables precise extraction of the threshold pump power. The calculated far-field profile shown in Fig.~\ref{L9_Cavity}(b) qualitatively agrees with measurements reported in \cite{Bonato2013} as well as our own experimental observations, Fig.~\ref{L9_Cavity}(c).  The calculation indicates a vertical collection efficiency of $\sim20\%$ for an L9 cavity, which may vary slightly for different cavity structures.

 \begin{figure}
 \includegraphics[width=6.6cm]{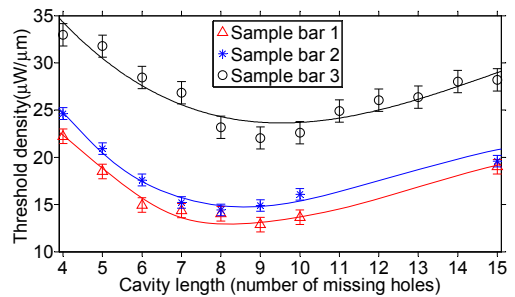}%
 \caption{\label{threshold} Measured threshold pump power density versus cavity length. Three sample bars ( red ``$\triangle$'', blue ``$\ast$'' and black ``$\circ$'') with the same PhC design but from different positions on the wafer are investigated. The solid lines are guides to the eye.}
 \end{figure}

Fig.~\ref{threshold} summarizes the measured variation of the laser threshold with cavity length for three sample bars with the same PhC design. The absolute number quoted for the threshold pump density is defined as $P_{exc}\Gamma_{exc}/L$, where $L$ is the cavity length. While bars 1 and 2 were located close to each other on the wafer and fabricated in the same run, bar 3 is more distant and is from another processing batch. Gain material variations and fabrication uncertainties are believed to account for the higher threshold of sample bar 3. The error bar indicates the largest variation of measured threshold between two repeated measurements. Lasing was not observed for L1, L2 and L3 cavities. Despite the difference in absolute gain level, all the sample bars show the same qualitative behaviour. From L4 to L8/L9, the threshold decreases monotonically with cavity length after which it increases. An optimum cavity length thus exists, corresponding to a cavity with 8 or 9 missing holes, for which the laser threshold density attains a minimum. This behavior is strikingly different from the case of conventional edge emitting lasers, where the threshold density decreases monotonously with length since the mirror losses decrease in inverse proportion to the cavity length.

In the measurements reported in Fig.~\ref{threshold}, we have taken a number of precautions to eliminate other causes for threshold variation. In particular, the pump profile is kept fixed with a diameter of $20 \ \mu \text{m}\  (\sim46a)$, which is much larger than the extent of all the investigated cavities from L1 to L15. This ensures a uniform illumination,  avoiding a systematic reduction of the pump efficiency with cavity length.  Heating induced by the optical pump beam leads to a red-shift of the laser wavelength. The use of an InP membrane already significantly improves the thermal properties compared to more conventional InGaAsP structures, due to the ten times larger thermal conductivity of InP, \cite{Xue2015}. To further reduce thermal effect, we employed pulsed pumping with a pulse width of 500 ns and a low duty cycle of 2\%. 

\begin{figure}
 \includegraphics[width=6.6cm]{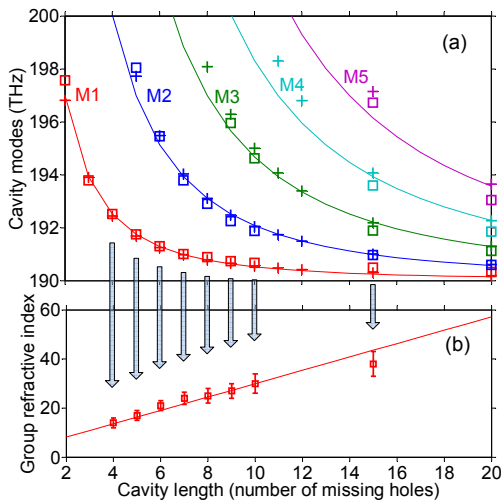}%
 \caption{\label{CavityModes}(a) Measured cavity mode frequencies (``square'' markers) versus cavity length. The solid lines are predictions by the analytical theory Eq.~(\ref{eq:two}). The ``cross'' markers are from the full numerical calculations. The fundamental mode, denoted M1 (red), is the dominating mode in all cases, and its threshold varies as shown in Fig.\ref{threshold}. (b) The measured group refractive index (markers) for the fundamental mode M1. The solid line is the calculation from Eq.~(\ref{eq:two}).}
 \end{figure}

The variation of the experimentally observed cavity mode frequencies with cavity length is shown by the ``square'' markers in Fig.~\ref{CavityModes}(a). As the cavity length increases, all modes shift to lower frequencies and the spacing between neighbouring modes decreases. Qualitatively similar behaviour was observed in \cite{Atlasov2009}. From the mode spacing and the known cavity length we can calculate the effective group refractive index. The extracted group index for the fundamental M1 mode, Fig.~\ref{CavityModes}(b), increases monotonously with cavity length, consistent with the mode approaching the BZ-edge. For the optimum cavity length, the group index is seen to be of the order of $\sim 20$. 

From experimental and theoretical studies of light propagation in PhC waveguides, it is already known that disorder leads to propagation losses that strongly increase in the slow-light region \cite{Hughes2005,Faolain2010}. Since the fundamental mode moves deeper into the slow-light region upon increasing the cavity length, it is perhaps not surprising that an optimum cavity length exists, as demonstrated in Fig.~\ref{threshold}. However, since the effective material gain per unit length also increases in the slow-light region \cite{Ek2014}, and the positive role of slow light on promoting laser oscillation in similar-type lasers was suggested \cite{Kiyota2006}, a more complete analysis is required.

We model the laser cavity as an effective Fabry-Perot resonator, but with the propagating wave to be a Bloch mode of the W1 waveguide. This was shown in \cite{Lalanne2008} to be a good approximation even for defect cavities as short as a few lattice periods. The effective length of the cavity is $L\cong(N+1)a$, where $N$ is the number of missing holes. By requiring that the Bloch mode-experiences a roundtrip phase change of an integer times $2\pi$ during a cavity roundtrip \cite{Bonato2013}, we find that the $M'th$ cavity mode is displaced by $\bigtriangleup k\cong M\pi/L$ from the BZ-edge located at $k_0=\pi/a$ \cite{Okano2010}. Close to this band-minimum we may expand the dispersion to second-order:

\begin{equation}
\omega(k)\cong\omega\left(\frac{\pi}{a}\right)+\frac{1}{2}\frac{d^2\omega}{dk^2}(\bigtriangleup k)^2.
\label{eq:one}
\end{equation}

The deviation of the mode frequency from the-BZ edge, $\bigtriangleup f$, and the corresponding light slow-down factor, $S=n_g^{phc}/n_g^b=v_g^b/v_g^{phc}$, thus become

\begin{eqnarray}
\bigtriangleup f=\frac{v_g^bl_2}{4}\frac{M^2}{L^2};\;
S=\frac{L}{Ml_2}.
\label{eq:two}
\end{eqnarray}

Here $v_g^x \ (n_g^x)$ is the group velocity (group refractive index) in the PhC cavity $(x=phc)$ and in the corresponding homogeneous material $(x=b)$. We also introduced the characteristic length scale $l_2=(\pi/v_g^b)(d^2\omega/dk^2)$, which is inversely proportional to an effective photon mass.

Solid lines in Fig.~\ref{CavityModes} show the frequencies for the first five cavity modes and the group index for the fundamental mode obtained from Eq.~(\ref{eq:two}) using $n_g^b=3.17, \  l_2=510\ \text{nm}$ and a band-edge frequency of 190 THz. The good agreement with experiment supports the validity of our approximate analytical model. Mode frequencies (``cross'' markers in Fig.~\ref{CavityModes}(a)) obtained from 3D finite-element simulations without adjustable dispersion parameters further confirm the model.

A rate equation for the photon density including the effect of slow light is derived by considering the incremental change of the cavity photon density, $N_p$, during one cavity roundtrip time, $\tau_R=2SL/v_g^b$. The  modal gain per unit length of a Bloch wave propagating in the structure is $g_{mod}=\Gamma S g_{mat}$, where $g_{mat}$ is the material gain coefficient for the homogeneous active medium and $\Gamma$ is the confinement factor \cite{Ek2014}. Neglecting spontaneous emission we find

\begin{equation}
\frac{dN_p}{dt}=v_g^b[\Gamma g_{mat}-(\alpha_{WG}+\alpha_{MIR})/S]N_p.
\label{eq:three}
\end{equation}

Here, $\alpha_{MIR}={\rm ln}[1/(R_1 R_2)]/(2L)$ and $\alpha_{WG}$ denote the mirror and waveguide loss. We see that the spatial gain enhancement is cancelled out due to the longer roundtrip time, which is basically a statement that the rate of gain per unit time is unaffected by slow-light \cite{Kiyota2006}. Fabrication induced disorder approximately scales as $\alpha_{WG}=S\alpha_1+S^2\alpha_2$ \cite{Hughes2005,Faolain2010}, where the first term, with coefficient $\alpha_1$, accounts for out-of-plane scattering and absorption losses and the second term, with coefficient $\alpha_2$, represents back-scattering. Since the latter component does not add coherently to the laser field we assume that it too acts as a loss term for the photon density of the considered laser mode. From the rate equation for the laser intensity we then find that the threshold gain becomes

\begin{equation}
\Gamma g_{th}=\alpha_1+\alpha_2 S+\frac{1}{2SL}{\rm ln}\left(\frac{1}{R_1R_2}\right)
\label{eq:four}
\end{equation}

\begin{figure}
 \includegraphics[width=7cm]{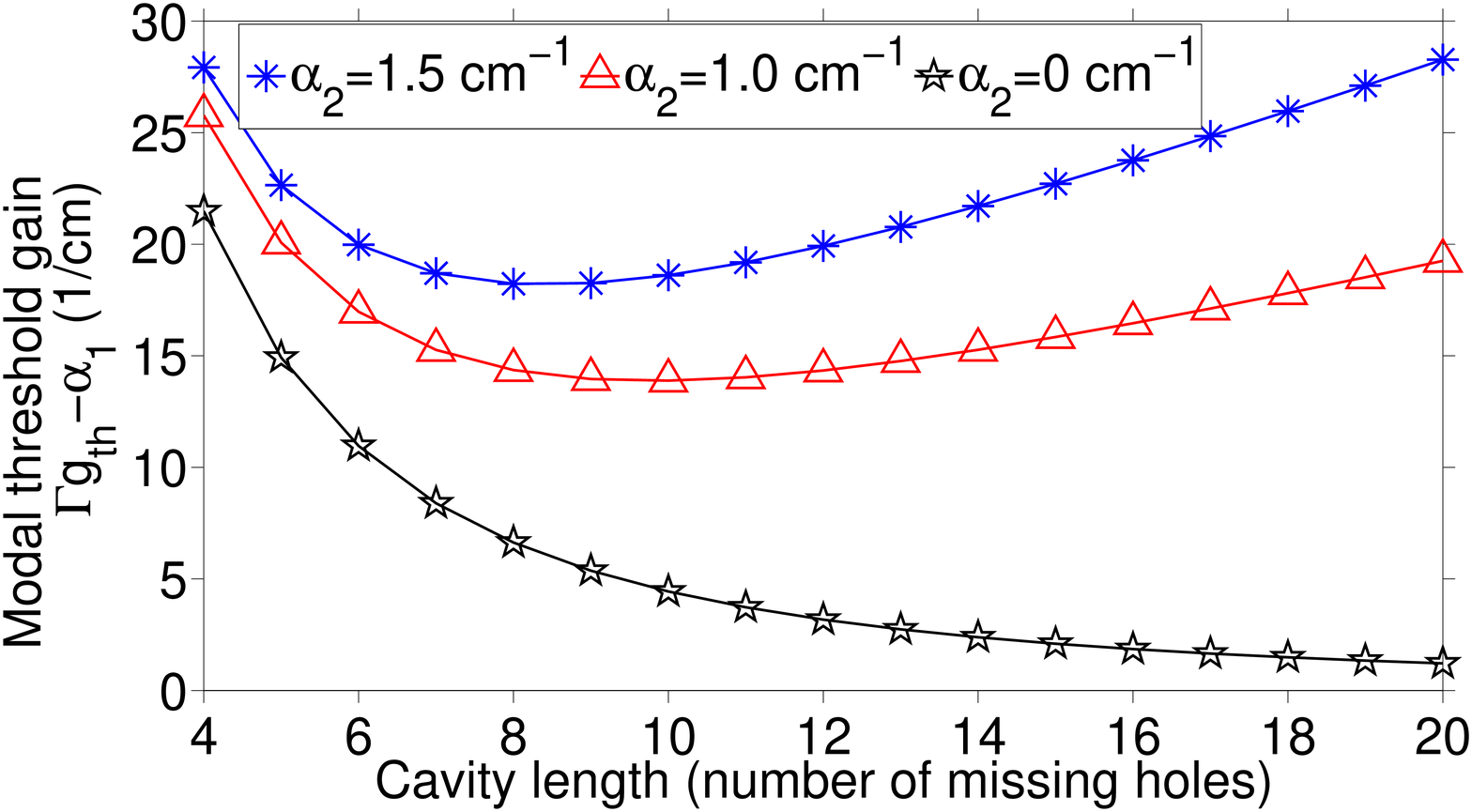}%
 \caption{\label{CalculatedThreshold}Calculated threshold gain, $\Gamma g_{th}-\alpha_1$ versus cavity length for different values of $\alpha_2$. $l_2= 510\ \text{nm}$, $R_1=R_2=0.98$.}
 \end{figure}

We see that the mirror loss is reduced in inverse proportion to the slow-down factor  $S$,  while disorder-induced losses originating from back-scatter increase linearly with the slow-down factor. Since $S$ itself scales linearly with $L$, short lasers are  dominated by mirror loss, while long lasers are governed by disorder-induced losses. Fig.~\ref{CalculatedThreshold} shows the calculated length dependence of the modal threshold gain, $\Gamma g_{th}-\alpha_1$, for the fundamental mode. For $\alpha_2=0\ \text{cm}^{-1}$, the threshold gain decreases monotonously with $L$, as for a conventional semiconductor laser, but for non-zero $\alpha_2$, an optimum cavity length is found, in good qualitative agreement with the experimental observations.  For the calculations presented here, we used  wavelength independent mirror reflectivities of $R_1=R_2=0.98$, but qualitatively similar results are obtained using a wavelength dependent reflectivity, as found, e.g., in \cite{Sauvan2005}. The value of $\alpha_2$ will depend on the actual level of fabrication disorder. For our structure, the standard deviation of the variation of the hole radius was measured to be of the order of 1 nm.

From these results  we infer that, in the presence of disorder, the quality factor of a PhC cavity is maximized for a specific cavity length. In order to further corroborate our finding we have addressed the problem using a different theoretical approach. In \cite{Savona2011} it was shown that the  loss of disordered waveguides can be efficiently computed by solving a generalized eigenvalue problem formulated in a basis of Bloch modes of the regular structure. Disorder leads to scattering among the Bloch modes and induces {\em extrinsic} losses of the guided modes, since these acquire Bloch mode components within the light cone. We apply the approach of \cite{Savona2011} to the case of a cavity, which is represented by a stepwise variation of the average permittivity along the propagation direction of the cavity, corresponding to an effective-index approach. Considering only a single band, the (generalized) eigenvalue problem can be formulated as 

\begin{equation}
\mathbf{D v_\beta}=\mathbf \omega_\beta \left(\mathbf{I}+\mathbf{V}\right)\mathbf{v_\beta}.
\label{eq:eigenvalueproblem}
\end{equation}

\begin{figure}
 \includegraphics[width=8.0 cm]{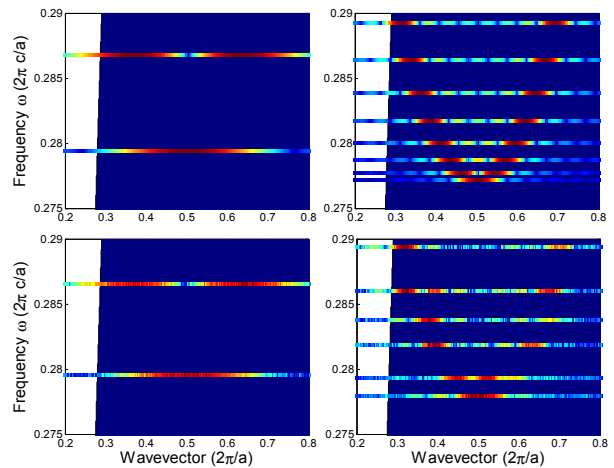}%
 \caption{\label{ModeSpectra} Calculated cavity mode spectra for L5 (left column) and L20 (right column) cavities in the absence of disorder (upper row) and with the inclusion of disorder (lower row). The Bloch mode distributions are shown in color-scale for the calculated mode eigenfrequencies (y-axis), with red (blue) denoting high (low) intensity. The white area indicates the ligt cone.}
 \end{figure}

Here, $\mathbf{D}$ is a diagonal matrix in the basis of Bloch modes, indexed by $k$, with  $\sqrt{D_{kk}}=\omega_k$ being the waveguide dispersion. Furthermore, $\mathbf{I}$ is the identity matrix and $\mathbf{V}=\mathbf{V^{cavity}}+\mathbf{V^{disorder}}$ is the scattering matrix coupling (hybridizing) the waveguide Bloch modes due to a determinsitic cavity ``potential" as well as a random``disorder" potential. The solution of Eq. (\ref{eq:eigenvalueproblem}) gives the eigenfrequencies of the cavity modes and the corresponding decomposition on Bloch modes. Fig.~\ref{ModeSpectra} shows the calculated mode spectra for L5 and L20 cavities with and without the inclusion of disorder. A similar graphical representation was used in \cite{Atlasov2009b}. In the absence of disorder, the fraction of the fundamental mode that is within the light cone decreases with cavity length, indicating that the cavity loss rate decreases \cite{Okano2010}. Upon including disorder, further scattering to the light cone takes place, reflecting the additional spatial frequency components introduced by the spatial disorder. 

\begin{figure}
 \includegraphics[width=6.6cm]{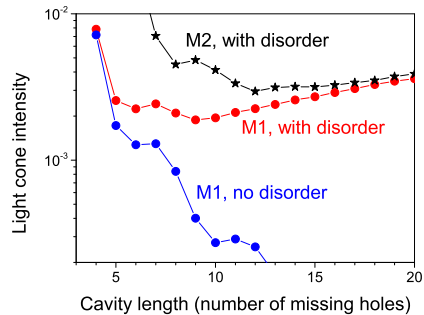}%
 \caption{\label{Fig:coneintensity} Calculated intensity within light cone versus cavity length for the fundamental mode M1, with and without disorder, and for the higher-order mode M2, with disorder.}
 \end{figure}

Fig.~\ref{Fig:coneintensity} shows the relative fraction of the mode spectrum within the light cone as a function of cavity length. An ensemble average over 500 realizations of the random scattering matrix $\mathbf{V^{disorder}}$ was performed. We  take this fraction as an approximate measure of the loss of that cavity mode. Quantitative results for the absolute loss rates can be obtained using the approach of \cite{Savona2011}, but is beyond the scope of this paper. While the results of course depend on the strength of the disorder potential and the effective index representation of the defect-cavity, the features displayed by Fig.~\ref{Fig:coneintensity} are general: In the presence of disorder, the light cone intensity increases, lending support to the assumption of our analytical model that disorder-induced back-scattering increases the cavity mode loss. Furthermore, the mode M1 is found always to be dominating. This is in accordance with our experimental findings, while the analytical result Eq. (\ref{eq:four}) suggests that higher-order modes become dominant deep into the slow-light region. 

In conclusion, we have experimentally shown that photonic crystal lasers display intriguing threshold characteristics, with a minimum threshold gain being attained at a specific cavity length.  An analytical model was derived that well accounts for the experimental results, suggesting that the observed threshold behavior is a result of the interplay between slow light and disorder-induced losses. These results are important for the understanding and further development of photonic crystal lasers. The results also demonstrate a promising platform for systematic experimental and theoretical investigations of disorder effects in active (amplifying) media and in cavities, which remain relatively unexplored compared to passive photonic crystal waveguides. For instance, a controlled amount of intentional disorder may be added \cite{Sapienza2010,Portalupi2011} and the relative strength of deterministic cavity modes versus random Anderson-localized modes \cite{Liu2014} might be controlled via the mirror reflectivities.

\begin{acknowledgments}
Financial support from the Danish Council for Independent Research (FTP 0602-02623b) and from the Villum Foundation via the NATEC Center and the Young Investigator Programme Project QUEENs are gratefully acknowledged.
\end{acknowledgments}



\bibliography{PhCLaser}

\end{document}